\documentclass[prb,showpacs,preprintnumbers,amsmath,amssymb]{revtex4}

\usepackage{epsfig}
\usepackage{graphicx}
\usepackage{dcolumn}
\usepackage{bm}

\newcommand{\EQ}{\begin{equation}}
\newcommand{\EN}{\end{equation}}
\newcommand{\ea}{\end{eqnarray}}
\newcommand{\ba}{\begin{eqnarray}}

\newcommand{\bear}{\begin{eqnarray}}
\newcommand{\ear}{\end{eqnarray}}

\begin{document}


\title{Global $SO(3)\times SO(3)\times U(1)$ symmetry of the Hubbard model on bipartite lattices}
\author{J. M. P. Carmelo$^{1}$, Stellan \"Ostlund$^{2}$, and M. J. Sampaio$^{1}$} 
\affiliation{$^1$GCEP-Centre of Physics, University of Minho, Campus Gualtar, P-4710-057 Braga, Portugal}
\affiliation{$^2$G\"oteborgs Universitet, Gothenburg 41296, Sweden}

\date{22 August 2009}


\begin{abstract}
In this paper the global symmetry of the Hubbard model on a bipartite lattice is
found to be larger than $SO (4)$. The model is one of the most studied many-particle quantum problems, yet
except in one dimension it has no exact solution, so that there remain many open questions about 
its properties. Symmetry plays an important role in physics and
often can be used to extract useful information on unsolved non-perturbative quantum problems.
Specifically, here it is found that for on-site interaction $U\neq 0$ the local 
$SU(2)\times SU(2) \times U(1)$ gauge symmetry of the 
Hubbard model on a bipartite lattice with $N_a^D$ sites and vanishing 
transfer integral $t=0$ can be lifted to a global 
$[SU(2)\times SU(2)\times U(1)]/Z_2^2=SO(3)\times SO(3)\times U(1)$ 
symmetry in the presence of the kinetic-energy hopping
term of the Hamiltonian with $t>0$. (Examples of a bipartite lattice are the $D$-dimensional 
cubic lattices of lattice constant $a$ and edge length $L=N_a\,a$ for which $D=1,2,3,...$ 
in the number $N_a^D$ of sites.) The generator of the 
new found hidden independent charge global $U(1)$ symmetry,
which is not related to the ordinary $U(1)$ gauge subgroup of electromagnetism,
is one half the rotated-electron number of singly-occupied sites 
operator. Although addition of chemical-potential and magnetic-field operator 
terms to the model Hamiltonian lowers its symmetry,
such terms commute with it. Therefore, its $4^{N_a^D}$ energy eigenstates refer
to representations of the new found global 
$[SU(2)\times SU(2)\times U(1)]/Z_2^2=SO(3)\times SO(3)\times U(1)$ symmetry.
Consistently, we find that for the Hubbard model on a bipartite lattice the number of independent 
representations of the group $SO(3)\times SO(3)\times U(1)$ equals
the Hilbert-space dimension $4^{N_a^D}$. It is confirmed elsewhere that the new found
symmetry has important physical consequences.
\end{abstract}
\pacs{02.20.Qs, 71.10.Fd}

\maketitle

\section{Introduction}

The Hubbard model on a bipartite lattice (for instance one-dimensional, square, 
cubic, and other $D$-dimensional cubic lattices) is the simplest realistic toy model for description of the 
electronic correlation effects in general many-particle problems with short-range 
interaction. It can be experimentally realized with unprecedented 
precision in systems of ultra-cold fermionic atoms on an optical lattice of variable geometry.
For the square and cubic lattices one may expect very detailed experimental results over a wide range of 
parameters to be available \cite{Zoller}. For instance, recently  
systems of ultra-cold fermionic atoms describing the Mott-Hubbard insulating 
phase of the the Hubbard model on a cubic lattice were studied \cite{cubic}. On the one dimensional
and square lattices the model has been widely used for describing the effects of 
correlations in several types of materials such as quasi-one-dimensional conductors 
\cite{Claessen,TTF-TCNQ} and high-$T_c$ superconductors \cite{ARPES-review,2D-MIT,duality}.

Unfortunately, most exact results and well-controlled approximations
for this model exist only in one dimension (1D) \cite{Lieb,Takahashi,Martins}.
Many open questions about its properties remain unsolved. One of the few
exact results, which refers to the model on any
bipartite lattice, is that for on-site interaction $U\neq 0$ it 
contains a global $SO(4)=[SU(2)\times SU(2)]/Z_2$ symmetry. It is 
associated with a spin $SU(2)$ symmetry and a 
charge $\eta$-spin $SU(2)$ symmetry \cite{Zhang}. 
We denote the $\eta$-spin (and spin) value 
of the energy eigenstates by $S_{\eta}$ (and $S_s$) and the corresponding 
projection by $S^z_{\eta}= -[N_a^D-N]/2$ (and $S^z_s= -[N_{\uparrow}-
N_{\downarrow}]/2$). Here $N_a^D$ denotes the number of lattice sites
and $N=N_{\uparrow}+N_{\downarrow}$ that of electrons.
Our notation $N_a^D$ is particularly appropriate to a $D$-dimensional 
cubic lattice where $D=1,2,3,...$ for the one-dimensional, square,
cubic ... lattice, respectively, $N_a$ is the number of sites in an
edge of length $L=N_a\,a$, and $a$ is the spacing.

In this paper we find that 
for $U\neq 0$ the local $SU(2)\times SU(2) \times U(1)$ gauge 
symmetry of the Hubbard model on a bipartite lattice with 
transfer integral $t=0$ \cite{U(1)-NL} can be lifted to a global 
$[SU(2)\times SU(2)\times U(1)]/Z_2^2=SO(3)\times SO(3)\times U(1)$
symmetry for the model with $t>0$. Indeed, the requirement of 
commutability with the $U/4t\neq 0$ interacting Hamiltonian replaces 
the $U=0$ global $O(4)/Z_2=[SO(4)\times Z_2]/Z_2$ symmetry by 
$SO(3)\times SO(3)\times U(1)=[SO(4)\times U(1)]/Z_2$ rather than $SO(4)$. Here, the factor 
$Z_2$ in $SO(4)\times Z_2$ refers to the particle-hole 
transformation on a single spin under which the interacting term is not
invariant \cite{U(1)-NL}. In $O(4)/Z_2$ and
$[SU(2)\times SU(2)\times U(1)]/Z_2^2=SO(3)\times SO(3)\times U(1)$ the factors 
$1/Z_2$ and $1/Z_2^2$ impose that
$[S_{\eta}+S_s]$ and $[S_{\eta}+S_s+S_c]$, respectively,
are integers. In the latter equations $S_c$ is the eigenvalue of
the generator of the new global $U(1)$ symmetry found in this paper.
Our results profit from those of Ref. \cite{Stein} and
reveal that such a symmetry
becomes explicit, provided that one describes the
problem in terms of rotated electrons. Those are generated 
by any of the unitary transformations considered in
that reference, which refer to $U/4t> 0$ values and can
be trivially extended to $U/4t\neq 0$ values. The global
symmetry found here refers to the latter $U/4t$ range.

The paper is organized as follows. The model and the electron - rotated-electron 
unitary transformations are the subjects of Section II. In Section III a global $SO(3)\times SO(3)\times U(1)$ 
symmetry is established for the Hubbard model on a bipartite lattice with $U/4t\neq 0$.
Finally, Section IV contains the concluding remarks.

\section{The model and a set useful electron - rotated-electron unitary transformations}

On a bipartite lattice with spacing $a$, $N_a^D\equiv [N_a]^D$ sites, $N_a$ even, 
$N_a/2$ odd, $L=N_a\,a$, and spatial dimension $D<N_a$ the Hubbard model is given by,
\begin{equation}
\hat{H} = \hat{T} + {U\over 2}[N_a^D-\hat{Q}] \, ;
\hspace{0.25cm}
\hat{T} = -t\sum_{\langle\vec{r}_j\vec{r}_{j'}\rangle}\sum_{\sigma =\uparrow
,\downarrow}[c_{\vec{r}_j\sigma}^{\dag}\,c_{\vec{r}_{j'}\sigma}+h.c.] \, ;
\hspace{0.25cm}
{\hat{Q}} = \sum_{j=1}^{N_a^D}\sum_{\sigma =\uparrow
,\downarrow}\,{\hat{n}}_{\vec{r}_j\sigma}\,(1- {\hat{n}}_{\vec{r}_j -\sigma}) \, .
\label{H}
\end{equation}
Here $\hat{T}$ is the kinetic-energy operator with first-neighbor transfer integral $t$,
which can be expressed in terms of the operators, 
\begin{eqnarray}
\hat{T}_0 & = & -\sum_{\langle\vec{r}_j\vec{r}_{j'}\rangle}\sum_{\sigma}[{\hat{n}}_{\vec{r}_{j},-\sigma}\,c_{\vec{r}_j,\sigma}^{\dag}\,
c_{\vec{r}_{j'},\sigma}\,{\hat{n}}_{\vec{r}_{j'},-\sigma} +
(1-{\hat{n}}_{\vec{r}_{j},-\sigma})\,c_{\vec{r}_j,\sigma}^{\dag}\,
c_{\vec{r}_{j'},\sigma}\,(1-{\hat{n}}_{\vec{r}_{j'},-\sigma})+h.c.] \, ,
\nonumber \\
\hat{T}_{+1} & = & -\sum_{\langle\vec{r}_j\vec{r}_{j'}\rangle}\sum_{\sigma}
{\hat{n}}_{\vec{r}_{j},-\sigma}\,c_{\vec{r}_j,\sigma}^{\dag}\,c_{\vec{r}_{j'},\sigma}\,(1-{\hat{n}}_{\vec{r}_{j'},-\sigma}) \, ,
\nonumber \\
\hat{T}_{-1} & = & -\sum_{\langle\vec{r}_j\vec{r}_{j'}\rangle}\sum_{\sigma}
(1-{\hat{n}}_{\vec{r}_{j},-\sigma})\,c_{\vec{r}_j,\sigma}^{\dag}\,
c_{\vec{r}_{j'},\sigma}\,{\hat{n}}_{\vec{r}_{j'},-\sigma} \, ,
\label{T-op}
\end{eqnarray}
as $\hat{T}= t\,[\hat{T}_0 + \hat{T}_{+1} + \hat{T}_{-1}]$. While the operator $\hat{T}_0$ does not change electron double 
occupancy, the operators $\hat{T}_{+1}$ and $\hat{T}_{-1}$ do it by $+1$ 
and $-1$, respectively. In the above equations
${\hat{n}}_{{\vec{r}}_j,\sigma} = c_{\vec{r}_j\sigma}^{\dag} c_{\vec{r}_j\sigma}$,
$\pm\sigma$ refer to opposite spin projections, and the operator ${\hat{Q}}$
counts the number of electron singly occupied sites. Hence
the operators,
\begin{equation}
{\hat{D}} = {1\over 2}[{\hat{N}}-{\hat{Q}}] \, ; \hspace{0.35cm} 
{\hat{D}}^h= {1\over 2}[{\hat{N}}^h-{\hat{Q}}] \, ; \hspace{0.35cm} 
{\hat{Q}}_{\uparrow} = {1\over 2}[{\hat{Q}}+({\hat{N}}_{\uparrow}-{\hat{N}}_{\downarrow})]
\, ; \hspace{0.35cm} 
{\hat{Q}}_{\downarrow} =  {1\over 2}[{\hat{Q}}-({\hat{N}}_{\uparrow}-{\hat{N}}_{\downarrow})] \, ,
\label{DDhQud}
\end{equation}
count the number of electron doubly
occupied sites, unoccupied sites, and spin $\sigma =\uparrow,\downarrow$ singly
occupied sites, respectively. Moreover, ${\hat{N}} = \sum_{\sigma}
{\hat{N}}_{\sigma}$ and ${\hat{N}}_{\sigma}=\sum_{j=1}^{N_a^D}
n_{{\vec{r}}_j,\sigma}$ where ${\hat{N}}^h = 
[2N_a^D - {\hat{N}}]$, ${\hat{N}}^h_{\uparrow}=
[N_a^D -{\hat{N}}_{\downarrow}]$, and ${\hat{N}}^h_{\downarrow}=
[N_a^D -{\hat{N}}_{\uparrow}]$. 

For simplicity let us consider that $U/4t>0$ and
let $\{\vert \Psi_{\infty}\rangle\}$ be a complete set of
$4^{N_a^D}$ energy eigenstates for $U/4t\rightarrow \infty$.
There is exactly one unitary operator
$\hat{V}=\hat{V}(U/4t)$ such that for the
value of $U/4t>0$ under consideration each of the $4^{N_a^D}$ states 
$\vert \Psi_{U/4t}\rangle=\hat{V}^{\dag}\vert \Psi_{\infty}\rangle$ 
is generated from the electronic vacuum by the same occupancy 
configurations of {\it rotated electrons} of creation operator
${\tilde{c}}_{\vec{r}_j\sigma}^{\dag}$
as the corresponding $U/4t\rightarrow \infty$ energy eigenstate in terms
of electrons. The rotated-electron creation and annihilation operators
are given by,
\begin{equation}
{\tilde{c}}_{\vec{r}_j,\sigma}^{\dag} =
{\hat{V}}^{\dag}\,c_{\vec{r}_j,\sigma}^{\dag}\,{\hat{V}}
\, ; \hspace{0.35cm}
{\tilde{c}}_{\vec{r}_j,\sigma} =
{\hat{V}}^{\dag}\,c_{\vec{r}_j,\sigma}\,{\hat{V}}
\, ; \hspace{0.35cm}
{\tilde{n}}_{\vec{r}_j,\sigma} = 
{\tilde{c}}_{\vec{r}_j,\sigma}^{\dag}\,{\tilde{c}}_{\vec{r}_j,\sigma} \, .
\label{rotated-operators}
\end{equation}

Rotated-electron single and double occupancy are good quantum
numbers for $U/4t> 0$ whereas for electrons
such occupancies become good
quantum numbers only for $U/4t\rightarrow\infty$. 
Therefore, $\hat{V}=\hat{V}(U/4t)$ becomes the unit 
operator in that limit. The unitary transformation 
associated with the operator $\hat{V}$ 
is of the type studied in Ref. \cite{Stein}.
There is one of such transformations for each 
choice of $U/4t\rightarrow \infty$ energy eigenstates. 
Similar results are obtained for $U/4t<0$. 

We introduce the operator ${\tilde{O}}={\hat{V}}^{\dag}\,{\hat{O}}\,{\hat{V}}$. It
has the same expression in terms of rotated-electron creation and annihilation
operators as ${\hat{O}}$ in terms of electron creation and annihilation operators.
Here ${\hat{V}}={\tilde{V}}$. Note that within our representation both the notations
referring to marks placed over letters being a caret ${\hat{W}}$ or a tilde ${\tilde{L}}$ 
denote operators. Such notations are useful for operators 
for which $W=L$ such as the general operators ${\hat{O}}$ and ${\tilde{O}}$. 
Indeed, then they imply the equivalent relations
${\tilde{O}}={\hat{V}}^{\dag}\,{\hat{O}}\,{\hat{V}}$
and ${\hat{O}}={\tilde{V}}\,{\tilde{O}}\,{\tilde{V}}^{\dag}$. (Here we
have used that ${\hat{V}}={\tilde{V}}$.) When ${\hat{O}}\neq {\tilde{O}}$
our convention is that in general the expression of the operator ${\hat{O}}$
in terms of electron creation and annihilation operators is
simpler than that of ${\tilde{O}}={\hat{V}}^{\dag}\,{\hat{O}}\,{\hat{V}}$
in terms such operators. This then implies that the expression of
${\tilde{O}}$ in terms of rotated-electron creation and annihilation operators is
simpler than that of ${\hat{O}}={\tilde{V}}\,{\tilde{O}}\,{\tilde{V}}^{\dag}$
in terms of the same rotated-electron operators. (An exception are the electron operators of Eq. (\ref{rotated-operators}), 
which denote by $c_{\vec{r}_j,\sigma}^{\dag}$ and
$c_{\vec{r}_j,\sigma}$  rather than by ${\hat{c}}_{\vec{r}_j,\sigma}^{\dag}$ and
${\hat{c}}_{\vec{r}_j,\sigma}$, respectively.)

Any operator ${\hat{O}}$ can be written as,
\begin{equation}
{\hat{O}} = {\hat{V}}\,{\tilde{O}}\,{\hat{V}}^{\dag}
= {\tilde{O}}+ [{\tilde{O}},\,{\hat{S}}\,] + {1\over
2}\,[[{\tilde{O}},\,{\hat{S}}\,],\,{\hat{S}}\,] + ... 
= {\tilde{V}}\,{\tilde{O}}\,{\tilde{V}}^{\dag}
= {\tilde{O}}+ [{\tilde{O}},\,{\tilde{S}}\,] + {1\over
2}\,[[{\tilde{O}},\,{\tilde{S}}\,],\,{\tilde{S}}\,] + ...\, ,
\label{OOr}
\end{equation}
where ${\hat{V}}^{\dag} = e^{{\hat{S}}}$, 
${\hat{V}} = e^{-{\hat{S}}}$, and
${\hat{S}}={\tilde{S}}$. That ${\hat{S}}$ and ${\hat{V}}$
have the same expression both in terms of
electron and rotated-electron creation and
annihilation operators justifies that ${\hat{O}} = {\hat{V}}\,{\tilde{O}}\,{\hat{V}}^{\dag} =
{\tilde{V}}\,{\tilde{O}}\,{\tilde{V}}^{\dag}$ in Eq. (\ref{OOr}).
Importantly, it follows from the results of Ref. \cite{Stein}
that for each electron - rotated-electron unitary transformation and corresponding
unitary operator ${\hat{V}}$ of the type considered in that reference the operator ${\hat{S}}$
has a uniquely defined expression. For any of such transformations
that unknown expression of ${\hat{S}}$ involves only the 
kinetic operators $\hat{T}_0$, $\hat{T}_{+1}$, and $\hat{T}_{-1}$
of Eq. (\ref{T-op}) and numerical $U/4t$ dependent coefficients.
For $U/4t\neq 0$ it can be expanded in a series of $t/U$. Importantly, the corresponding first-order
term has a universal form for all electron - rotated-electron unitary transformations
of the above-mentioned type, which reads \cite{Stein},
\begin{equation}
{\hat{S}} = -{t\over U}\,\left[\hat{T}_{+1} -\hat{T}_{-1}\right] 
+ {\cal{O}} (t^2/U^2) = {\tilde{S}} = -{t\over U}\,\left[\tilde{T}_{+1} -\tilde{T}_{-1}\right] 
+ {\cal{O}} (t^2/U^2) \, .
\label{S-largeU}
\end{equation}
(The form of our relation ${\hat{V}}^{\dag} = e^{{\hat{S}}}$ justifies the extra minus sign in the
${\hat{S}}$ and ${\tilde{S}}$ expressions given here, relative to those of Ref.
\cite{Stein}.)

Furthermore, for any unitary operator 
${\hat{V}}$ of the above type, $-{\hat{S}}$ can be written as 
$-{\hat{S}}={\hat{S}}(\infty)+\Delta {\hat{S}}$. Here ${\hat{S}}(\infty)$
corresponds to the operator $S(l)$ for $l=\infty$ defined in Eq. (61) 
of Ref. \cite{Stein} and $\Delta {\hat{S}}$ has the general form
provided in Eq. (64) of that reference. For each specific transformation
and corresponding choice of $U/4t\rightarrow\infty$ energy
eigenstates there is exactly one choice for the 
numbers $D^{(k)}(\bf{m})$ in that equation. ($k=1,2,...$ 
refers to the number of rotated-electron doubly occupied sites.)

Since $\hat{V}$ is unitary, the operators
${\tilde{c}}_{\vec{r}_j\sigma}^{\dag}$ and ${\tilde{c}}_{\vec{r}_j\sigma}$
have the same anticommutation relations as 
$c_{\vec{r}_j\sigma}^{\dag}$ and $c_{\vec{r}_j\sigma}$. 
The $\sigma$ electron number operator ${\hat{N}}_{\sigma}=\sum_{j=1}^{N_a^D}
{\hat{n}}_{\vec{r}_j,\sigma}$
equals the corresponding $\sigma$ rotated-electron number
operator $\tilde{N}_{\sigma} =\sum_{j=1}^{N_a^D}{\tilde{n}}_{\vec{r}_j,\sigma}$.
As a result, it remains invariant under $\hat{V}$, so that
$[\hat{N}_{\sigma},\hat{V}]=[\hat{N}_{\sigma},\hat{S}]=0$.
(See equation (\ref{OOr}) such that $[\tilde{N}_{\sigma},\hat{S}]=0$
for $\hat{O}=\hat{N}_{\sigma}$ and $\tilde{O}=\tilde{N}_{\sigma}$.)

\section{The global $SO(3)\times SO(3)\times U(1)$ symmetry for $U/4t\neq 0$}

\subsection{Global symmetry of the Hubbard model on a general bipartite lattice}

The local $SU(2)\times SU(2) \times U(1)$ gauge
symmetry of the Hamiltonian (\ref{H}) for $U/4t\rightarrow\pm\infty$ considered in Ref. \cite{U(1)-NL}
becomes for finite $\vert U/4t\vert > 0$ values a group of permissible unitary 
transformations. It is such that the corresponding local $U(1)$ 
canonical transformation is not the ordinary $U(1)$ gauge 
subgroup of electromagnetism. Instead it is a ``nonlinear" 
transformation \cite{U(1)-NL}. 
Following the unitary character of $\hat{V}=\tilde{V}$,
one can either consider that,
\begin{equation}
{\hat{H}} = {\hat{V}}\,{\tilde{H}}\,{\hat{V}}^{\dag}
= {\tilde{V}}\,{\tilde{H}}\,{\tilde{V}}^{\dag} = {\tilde{H}} + [{\tilde{H}},\,{\tilde{S}}\,] + {1\over
2}\,[[{\tilde{H}},\,{\tilde{S}}\,],\,{\tilde{S}}\,] + ... \, ,
\label{HHr}
\end{equation}
is the Hubbard model written in terms of rotated-electron
operators or another Hamiltonian with an involved expression
and whose operators
${\tilde{c}}_{\vec{r}_j\sigma}^{\dag}$ and 
${\tilde{c}}_{\vec{r}_j\sigma}$ refer to electrons. 
According to Ref. \cite{Stein}, the latter rotated
Hamiltonian is built up by use of the conservation of singly 
occupancy $2S_c=\langle {\tilde{Q}}\rangle$ 
by eliminating terms in the $t>0$
Hubbard Hamiltonian. That is done so that $S_c$ is an 
eigenvalue of the following one-half rotated-electron 
singly-occupancy number operator associated with the
operator ${\hat{S}}_c \equiv  {\hat{Q}}/2$,
\begin{equation}
{\tilde{S}}_c \equiv  {1\over 2}\,{\hat{V}}^{\dag}\,{\hat{Q}}\,{\hat{V}}
= {1\over 2}\,{\tilde{Q}} = {1\over 2}\sum_{j=1}^{N_a^D}\sum_{\sigma =\uparrow
,\downarrow}\,{\tilde{n}}_{\vec{r}_j\sigma}\,(1- {\tilde{n}}_{\vec{r}_j -\sigma})  \, .
\label{Or-ope}
\end{equation}
Here ${\tilde{n}}_{\vec{r}_j,\sigma}=
{\hat{V}}^{\dag}\,{\hat{n}}_{\vec{r}_j,\sigma}\,{\hat{V}}=
{\tilde{c}}_{\vec{r}_j\sigma}^{\dag}{\tilde{c}}_{\vec{r}_j\sigma}$
is the operator given in Eq. (\ref{rotated-operators}).
According to the studies of Ref. \cite{Stein}, this can be done to all 
orders of $t/U$ provided that $U/4t\neq 0$. In the context of Ref. 
\cite{Stellan-06}, this is equivalent to compute rotated
``quasicharge'' fermions whose number exactly equals 
$[N_a^D-2S_c]$. 

The ``rotated'' Hamiltonian ${\tilde{H}}={\hat{V}}^{\dag}\,{\hat{H}}\,{\hat{V}}$ commutes with the six generators of the
$SO(4)$ symmetry. Thus the Hubbard model ${\hat{H}}$ commutes with both such generators and
corresponding six other operators with the same expressions when written in terms of
rotated-electron operators. Consistently with Eq. (\ref{OOr}), this just means
that the six generators of the $\eta$-spin and spin algebras commute with ${\hat{V}}$. 
To reach this result we have profited from the expression of the operator ${\hat{S}}$ only involving 
the three kinetic operators given in Eq. (\ref{T-op}). We have then calculated the following 
commutators,
\begin{equation}
[{\hat{S}}_{\alpha}^z,\hat{T}_l] = [{\hat{S }}_{\alpha}^{\dagger},\hat{T}_l] = [{\hat{S }}_{\alpha},\hat{T}_l] =0 \, ; \hspace{0.25cm} \alpha = \eta , s
\, , \hspace{0.15cm} l=0,\pm 1 \, .
\label{S-T}
\end{equation}
Although the algebra involved in their derivation is cumbersome, it is straightforward. Therefore, we omit here the corresponding details. 
The vanishing of the commutators (\ref{S-T}) implies that the six generators of the $\eta$-spin and spin algebras commute with ${\hat{V}}$,
\begin{equation}
[{\hat{S}}_{\alpha}^z,{\hat{V}}] = [{\hat{S }}_{\alpha}^{\dagger},{\hat{V}}] = [{\hat{S }}_{\alpha},{\hat{V}}] =0 \, ; \hspace{0.25cm} \alpha = \eta , s \, .
\label{S-V-dag}
\end{equation}
This confirms that for such six operators all operator terms on the 
right-hand side of Eq. (\ref{OOr}) containing commutators vanish so that ${\hat{O}} = {\tilde{O}}$
for ${\hat{O}}$ being any of such operators. 
Hence they have the same expression in terms of electron and rotated-electron operators and read,
\begin{eqnarray}
{\hat{S}}_{\eta}^z & = & -{1\over 2}[N_a^D-\hat{N}] = -{1\over 2}[N_a^D-\tilde{N}] \, ; \hspace{0.35cm}
{\hat{S}}_s^z = -{1\over 2}[{\hat{N}}_{\uparrow}- {\hat{N}}_{\downarrow}] = -{1\over 2}[{\tilde{N}}_{\uparrow}- {\tilde{N}}_{\downarrow}] \, ,
\nonumber \\
{\hat{S }}_{\eta}^{\dagger} & = & \sum_{j=1}^{N_a^D}e^{i\vec{\pi}\cdot\vec{r}_j}\,c_{\vec{r}_j\downarrow}^{\dagger}\,
c_{\vec{r}_j\uparrow}^{\dagger} =\sum_{j=1}^{N_a^D}e^{i\vec{\pi}\cdot\vec{r}_j}\,{\tilde{c}}_{\vec{r}_j\downarrow}^{\dagger}\,
{\tilde{c}}_{\vec{r}_j\uparrow}^{\dagger} \, ; \hspace{0.35cm}
{\hat{S}}_{\eta} = \sum_{j=1}^{N_a^D}e^{-i\vec{\pi}\cdot\vec{r}_j}\,c_{\vec{r}_j\uparrow}\,c_{\vec{r}_j\downarrow} 
=\sum_{j=1}^{N_a^D}e^{-i\vec{\pi}\cdot\vec{r}_j}\,{\tilde{c}}_{\vec{r}_j\uparrow}\,{\tilde{c}}_{\vec{r}_j\downarrow} \, ,
\nonumber \\
{\hat{S}}_s^{\dagger} & = &
\sum_{j=1}^{N_a^D}\,c_{\vec{r}_j\downarrow}^{\dagger}\,c_{\vec{r}_j\uparrow} =
\sum_{j=1}^{N_a^D}\,{\tilde{c}}_{\vec{r}_j\downarrow}^{\dagger}\,{\tilde{c}}_{\vec{r}_j\uparrow} 
\, ; \hspace{0.35cm}
{\hat{S}}_s = \sum_{j=1}^{N_a^D}c_{\vec{r}_j\uparrow}^{\dagger}\,
c_{\vec{r}_j\downarrow}=\sum_{j=1}^{N_a^D}{\tilde{c}}_{\vec{r}_j\uparrow}^{\dagger}\,
{\tilde{c}}_{j,\,\downarrow} \, , \label{Scs}
\end{eqnarray}
where the vector $\vec{\pi}$ has Cartesian components $\vec{\pi}=[\pi,\pi,...]$. For instance,
for the model on the 1D, square, and cubic lattices those read $\pi$, $[\pi,\pi]$, and $[\pi,\pi,\pi]$, respectively.

In addition, we have evaluated the commutators of the three components
of the momentum operator $\hat{\vec{P}}$ with the three operators of Eq. (\ref{T-op}).
Again all such commutators vanish, so that the momentum operator
commutes with ${\hat{V}}$. Use of Eq. (\ref{OOr}) then implies 
that such an operator reads, 
\begin{equation}
\hat{{\vec{P}}}  = \sum_{\sigma=\uparrow ,\,\downarrow }\sum_{\vec{k}}\,\vec{k}\,
c_{\vec{k},\,\sigma }^{\dag }\,c_{\vec{k},\,\sigma } =
\sum_{\sigma=\uparrow ,\,\downarrow }\sum_{\vec{k}}\,\vec{k}\,
{\tilde{c}}_{\vec{k},\,\sigma }^{\dag }\,{\tilde{c}}_{\vec{k},\,\sigma } \, .
\label{P-invariant}
\end{equation}
Again all operator terms on the right-hand side of Eq. (\ref{OOr}) containing commutators vanish
for ${\hat{O}}$ being any of the three operator components of $\hat{\vec{P}}$, 
so that $\hat{{\vec{P}}} = \tilde{{\vec{P}}}$.

According to the studies of Ref. \cite{U(1)-NL}, the $SU(2)\times SU(2) \times U(1)$ Lie group and
its local generators can be represented by the $4\times 4$ on-site matrix 
$x_{\vec{r}_j}$ provided in Eq. (7) of that reference and matrices 
$o_{\vec{r}_j}$ appropriate to these generators. Their entries are given 
through polynomials of electron operators of the general form
${\hat{X}}_{\vec{r}_j} = \sum_{l,l'} x_{\vec{r}_j,l,l'}\, {\hat{m}}_{\vec{r}_j,l',l} 
\equiv {\rm Tr}\, (x_{\vec{r}_j}\, {\hat{m}}_{\vec{r}_j})$ and
${\hat{O}}_{\vec{r}_j} = \sum_{l,l'} o_{\vec{r}_j,l,l'}\, {\hat{m}}_{\vec{r}_j,l',l} 
\equiv {\rm Tr}\, (o_{\vec{r}_j}\, {\hat{m}}_{\vec{r}_j})$,
respectively. Here the operator matrix ${\hat{m}}_{\vec{r}_j}$ has
the same form as the operator matrix 
${\tilde{m}}_{\vec{r}_j}={\hat{V}}^{\dag}\,{\hat{m}}_{\vec{r}_j}\, {\hat{V}}$, 
but with the rotated-electron operators
replaced by electron operators. The operator matrix ${\tilde{m}}_{\vec{r}_j}$
plays an important role in our studies. It reads,
\begin{widetext}
{\bear {\tilde{m}}_{\vec{r}_j} =\left[
\begin{array}{cccc}
1 - {\tilde{n}}_{\vec{r}_j,\uparrow} - {\tilde{n}}_{\vec{r}_j,\downarrow} + {\tilde{n}}_{\vec{r}_j,\uparrow}\, {\tilde{n}}_{\vec{r}_j,\downarrow}
& {\tilde{c}}_{\vec{r}_j,\downarrow}\, {\tilde{c}}_{\vec{r}_j,\uparrow} &
(1-{\tilde{n}}_{\vec{r}_j,\downarrow})\, {\tilde{c}}_{\vec{r}_j,\uparrow} & (1-{\tilde{n}}_{\vec{r}_j,\uparrow})\, {\tilde{c}}_{\vec{r}_j,\downarrow} \\
{\tilde{c}}^{\dag}_{\vec{r}_j,\uparrow}\, {\tilde{c}}^{\dag}_{\vec{r}_j,\downarrow} & {\tilde{n}}_{\vec{r}_j,\uparrow}\, {\tilde{n}}_{\vec{r}_j,\downarrow} &
- {\tilde{c}}^{\dag}_{\vec{r}_j,\downarrow}\, {\tilde{n}}_{\vec{r}_j,\uparrow} & {\tilde{c}}^{\dag}_{\vec{r}_j,\uparrow}\, {\tilde{n}}_{\vec{r}_j,\downarrow} \\
{\tilde{c}}^{\dag}_{\vec{r}_j,\uparrow}\, (1-{\tilde{n}}_{\vec{r}_j,\downarrow}) & -{\tilde{n}}_{\vec{r}_j,\uparrow}\, {\tilde{c}}^{\dag}_{\vec{r}_j,\downarrow} &
{\tilde{n}}_{\vec{r}_j,\uparrow}\, (1-{\tilde{n}}_{\vec{r}_j,\downarrow}) & {\tilde{c}}^{\dag}_{\vec{r}_j,\uparrow}\, {\tilde{c}}_{\vec{r}_j,\downarrow} \\
{\tilde{c}}^{\dag}_{\vec{r}_j,\downarrow}\, (1-{\tilde{n}}_{\vec{r}_j,\uparrow}) & {\tilde{n}}_{\vec{r}_j,\downarrow}\, {\tilde{c}}_{\vec{r}_j,\uparrow} &
{\tilde{c}}^{\dag}_{\vec{r}_j,\downarrow}\, {\tilde{c}}_{\vec{r}_j,\uparrow} & {\tilde{n}}_{\vec{r}_j,\downarrow}\, (1-{\tilde{n}}_{\vec{r}_j,\uparrow}) 
\label{tilde-MS}
\end{array}\right] \, .
\ear}
\end{widetext}
As described in Ref. \cite{U(1)-NL}
for the polynomial ${\hat{O}}_{\vec{r}_j}$, one can as well introduce 
a general polynomial operator ${\tilde{O}}_{\vec{r}_j}$
of rotated-electron operators of the general form,
\begin{equation}
{\tilde{O}}_{\vec{r}_j} = \sum_{l,l'} o_{\vec{r}_j,l,l'}\, {\tilde{m}}_{\vec{r}_j,l',l} 
\equiv {\rm Tr}\, (o_{\vec{r}_j}\, {\tilde{m}}_{\vec{r}_j}) \, .
\label{tilde-X-r}
\end{equation}

Lifting the local $\eta$-spin and spin $SU(2)\times SU(2)$ gauge 
symmetry of the Hubbard model on a bipartite lattice for
$U/4t=\pm\infty$ to a global $[SU(2)\times SU(2)]/Z_2=SO(4)$ symmetry 
of that model for $U/4t\neq 0$ is simply 
accomplished by summing over the $N_a^D$
sites the six local generators ${\hat{O}}_{\vec{r}_j}$
of the $SU(2)\times SU(2)$ sub-group of the
$SU(2)\times SU(2) \times U(1)$ Lie group. 
It follows from the equalities of Eq. (\ref{Scs})
that the six generators given in that equation
can be represented by polynomials of electron
and rotated-electron operators of the same form,
$\sum_{j=1}^{N_a^D} {\hat{O}}_{\vec{r}_j} = \sum_{j=1}^{N_a^D} {\tilde{O}}_{\vec{r}_j}$.
This holds in spite of except for $U/4t\rightarrow\pm\infty$ the corresponding
local generators ${\hat{O}}_{\vec{r}_j}$ and ${\tilde{O}}_{\vec{r}_j}$ being different operators,
${\hat{O}}_{\vec{r}_j} \neq {\tilde{O}}_{\vec{r}_j}$. Indeed, the
local generators ${\hat{O}}_{\vec{r}_j}$ do not in general commute with the
unitary operator $\hat{V}$. This follows from ${\hat{m}}_{\vec{r}_j,l',l} \neq {\tilde{m}}_{\vec{r}_j,l',l}$,
where ${\hat{m}}_{\vec{r}_j,l',l}$ and ${\tilde{m}}_{\vec{r}_j,l',l}$ appear
in the expressions ${\hat{O}}_{\vec{r}_j} = \sum_{l,l'} o_{\vec{r}_j,l,l'}\, {\hat{m}}_{\vec{r}_j,l',l} 
\equiv {\rm Tr}\, (o_{\vec{r}_j}\, {\hat{m}}_{\vec{r}_j})$ and (\ref{tilde-X-r}) of ${\tilde{O}}_{\vec{r}_j}$, respectively.
However, the matrix $o_{\vec{r}_j}$ appearing in these two expressions is the same.
For the six local generators associated with the 
generators (\ref{Scs}) of the global $SO(4)$ symmetry it reads,  
{\bear o_{\vec{r}_j} =\left[
\begin{array}{cccc}
-1/2 & 0 & 0 & 0 \\
0 & 1/2 & 0 & 0 \\
0 & 0 & 0 & 0 \\
0 & 0 & 0 & 0 
\label{x-S-etaz}
\end{array}\right] \, ; \hspace{0.15cm} 
o_{\vec{r}_j} =\left[
\begin{array}{cccc}
0 & 0 & 0 & 0 \\
0 & 0 & 0 & 0 \\
0 & 0 & -1/2 & 0 \\
0 & 0 & 0 & 1/2
\label{x-S-sz}
\end{array}\right] 
\ear}for the $\eta$-spin and spin diagonal generators and
{\bear o_{\vec{r}_j} =\left[
\begin{array}{cccc}
0 & -e^{i\vec{\pi}\cdot\vec{r}_j} & 0 & 0 \\
0 & 0 & 0 & 0 \\
0 & 0 & 0 & 0 \\
0 & 0 & 0 & 0 
\end{array}\right] \, ; \hspace{0.15cm} 
o_{\vec{r}_j} =\left[
\begin{array}{cccc}
0 & 0 & 0 & 0 \\
0 & 0 & 0 & 0 \\
0 & 0 & 0 & 0 \\
0 & 0 & 1 & 0
\label{x-S-s+}
\end{array}\right] 
\ear}plus their two hermitian conjugates 
for the $\eta$-spin and spin off-diagonal generators. 

Now for the ``rotated'' Hamiltonian ${\tilde{H}} = {\hat{V}}^{\dag}\,{\hat{H}}\,{\hat{V}}= {\tilde{V}}^{\dag}\,{\hat{H}}\,{\tilde{V}}$ 
a local $SU(2)\times SU(2) \times U(1)$ gauge symmetry 
occurs for $U/4t\rightarrow\pm\infty$ as well. Alike the original Hamiltonian, ${\tilde{H}}$ has a
global $SO(4)$ symmetry whose generators are obtained as above. In addition, 
a similar procedure can be used to lift the local $U(1)$ gauge symmetry
to a global symmetry of the ``rotated''  Hamiltonian for $t>0$ and $U/4t\neq 0$. 
Indeed, through the polynomial of rotated-electron operators 
given in Eq. (\ref{tilde-X-r}), the local generator of the ``nonlinear" 
local $U(1)$ gauge symmetry can be represented by 
a $4\times 4$ matrix given by,{\bear 
o_{\vec{r}_j} =\left[
\begin{array}{cccc}
0 & 0 & 0 & 0 \\
0 & 0 & 0 & 0 \\
0 & 0 & 1/2 & 0 \\
0 & 0 & 0 & 1/2
\label{x-S-U1-p}
\end{array}\right] \, .
\ear}This local generator refers to rotated-electron single occupancy $2S_c$. 
The use of such a matrix $o_{\vec{r}_j}$ in the
polynomial ${\tilde{O}}_{\vec{r}_j}$ of Eq. (\ref{tilde-X-r})
leads for $U/4t\neq0$ to a sum of polynomials
$\sum_{j=1}^{N_a^D} {\tilde{O}}_{\vec{r}_j}$. It exactly equals 
expression (\ref{Or-ope}) of the generator of the 
global $U(1)$ symmetry whose eigenvalue $S_c$
is one half the number of rotated-electron singly occupied sites  
$2S_c$. The trivially related 
operator ${\tilde{S}}_c^h\equiv [{\tilde{D}}+{\tilde{D}}^h]/2$
of eigenvalue $S_c^h=[N_a^D/2-S_c]$ can also
generate such a global symmetry of the ``rotated'' Hamiltonian
${\tilde{H}} = {\hat{V}}^{\dag}\,{\hat{H}}\,{\hat{V}}= {\tilde{V}}^{\dag}\,{\hat{H}}\,{\tilde{V}}$. 
When written in terms of local polynomials as $\sum_{j=1}^{N_a^D} {\tilde{O}}_{\vec{r}_j}$,
its corresponding matrix $o_{\vec{r}_j}$ reads,
{\bear o_{\vec{r}_j} =\left[
\begin{array}{cccc}
1/2 & 0 & 0 & 0 \\
0 & 1/2 & 0 & 0 \\
0 & 0 & 0 & 0 \\
0 & 0 & 0 & 0 
\label{x-S-U1-h}
\end{array}\right] \, .
\ear}This generator refers to rotated-electron double occupancy plus non occupancy $[2N_a^D-2S_c]$. However,
$[2N_a^D-2S_c]$ and rotated-electron single occupancy $2S_c$ are not independent. Hence
the operators ${\tilde{S}}_c \equiv  {\tilde{Q}}/2$ of Eq. (\ref{Or-ope}) associated with the matrix $o_{\vec{r}_j}$
of Eq. (\ref{x-S-U1-p}) and ${\tilde{S}}_c^h\equiv [{\tilde{D}}+{\tilde{D}}^h]/2$ associated with the matrix $o_{\vec{r}_j}$
of Eq. (\ref{x-S-U1-h}) refer to two alternative representations of the generator of the global $U(1)$ symmetry
of the ``rotated'' Hamiltonian under consideration.
 
The main point is that a global $U(1)$ symmetry in the
``rotated'' Hamiltonian ${\tilde{H}} = {\hat{V}}^{\dag}\,{\hat{H}}\,{\hat{V}}= {\tilde{V}}^{\dag}\,{\hat{H}}\,{\tilde{V}}$ 
for $t>0$ and $U/4t\neq 0$ must
also be a global $U(1)$ symmetry, which is {\it hidden}
in the original model ${\hat{H}}={\hat{V}}\,{\tilde{H}}\,{\hat{V}}^{\dag}={\tilde{V}}\,{\tilde{H}}\,{\tilde{V}}^{\dag}$ of Eq. (\ref{HHr}). 
Indeed, for the latter original model the generator (\ref{Or-ope}) refers
to one half the number of rotated electrons rather than electrons. And in contrast to the
six generators (\ref{Scs}) of the global $SO(4)$ symmetry, the number of rotated electrons
operator does not commute with the unitary operator $\hat{V}$. Other related operators 
${\tilde{D}}$, ${\tilde{D}}^h$, and ${\tilde{Q}}_{\sigma}$,
which for $U/4t\neq 0$ also commute with the Hamiltonian (\ref{H}) yet do not
commute with $\hat{V}$, are obtained by rotating the number operators 
${\hat{D}}$, ${\hat{D}}^h$, and ${\hat{Q}}_{\sigma}$,
respectively, given in Eq. (\ref{DDhQud}). For the $N_a^D$-site problem only for rotated
electrons does single and double occupancy remain
good quantum numbers for finite $\vert U/4t\vert> 0$,
whereas for electrons single and double occupancy are
conserved only for $\vert U/4t\vert\rightarrow\infty$. 
This is why the generator (\ref{Or-ope})
does not commute with ${\hat{V}}$. 

Since $N_a^D$ is even, both $[S_{\eta}+S_s^z]$
and $[S_{\eta}^z+S_s^z]$ are integers. Their relation to $S_c$ is such that
$2S_c$ and $[N_a^D-2S_c]$ give the number of spin-$1/2$ spins of
the rotated electrons that singly occupy sites and the number of $\eta$-spin-$1/2$ "$\eta$-spins",
respectively. The former number equals that of rotated-electron singly occupied
sites and the latter number that of rotated-electron doubly occupied sites (down
$\eta$-spins) plus rotated-electron unoccupied sites (up $\eta$-spins), respectively. Therefore,   
$[S_{\eta}^z+S_s^z+S_c]$ must also be an integer. This justifies why 
for $U/4t\neq 0$ the global symmetry of the model
(\ref{H}) on a bipartite lattice is that of the group 
$[SU(2)\times SU(2) \times U(1)]/Z_2^2=SO(3)\times SO(3)\times U(1)$
rather than $SU(2)\times SU(2) \times U(1)$.
The global $U(1)$ symmetry remained hidden because
in contrast to the six generators (\ref{Scs}), one has that
the generator ${\tilde{S}}_c=\sum_{j=1}^{N_a^D} {\tilde{O}}_{\vec{r}_j}$
is except for $\vert U/4t\vert\rightarrow\infty$ different from the
operator $\sum_{j=1}^{N_a^D} {\hat{O}}_{\vec{r}_j}$.  
(Note that the matrix $o_{\vec{r}_j}$ is given by Eq. (\ref{x-S-U1-p}) 
both in the ${\tilde{O}}_{\vec{r}_j}$
and ${\hat{O}}_{\vec{r}_j}$ expressions.) Indeed, when 
written in terms of electron creation and annihilation operators the expression 
of the generator ${\tilde{S}}_c$ is for $\vert U/4t\vert$ finite involved, consisting of
an infinite number of operator terms,
\begin{equation}
{\tilde{S}}_c = \sum_{j=1}^{N_a^D} {\hat{V}}^{\dag}\,{\hat{O}}_{\vec{r}_j}\,{\hat{V}}
= \sum_{j=1}^{N_a^D}\left({\hat{O}}_{\vec{r}_j} + [{\hat{O}}_{\vec{r}_j},\,{\hat{S}}^{\dag}\,] + {1\over
2}\,[[{\hat{O}}_{\vec{r}_j},\,{\hat{S}}^{\dag}\,],\,{\hat{S}}^{\dag}\,] + ...\right) \, ,
\label{S-c-compli}
\end{equation}
rather than merely by $\sum_{j=1}^{N_a^D} {\hat{O}}_{\vec{r}_j}$. Only for
$\vert U/4t\vert\rightarrow\infty$ one has that the commutator
$[{\hat{O}}_{\vec{r}_j},\,{\hat{S}}^{\dag}\,]=0$ vanishes, so that 
${\tilde{S}}_c ={\hat{S}}_c =\sum_{j=1}^{N_a^D} {\hat{O}}_{\vec{r}_j}$.

\subsection{Consistency between the global symmetry and the Hilbert space dimension}

Addition of chemical-potential and magnetic-field operator 
terms to the Hamiltonian (\ref{H}) lowers its symmetry. 
However, such terms commute with it. Therefore, the global symmetry being
$[SU(2)\times SU(2)\times U(1)]/Z_2^2=SO(3)\times SO(3)\times U(1)$ 
implies that the set of independent rotated-electron occupancy configurations
that generate the model energy eigenstates   
generate state representations of that global symmetry for all values of the 
electronic density $n$ and spin density $m$. It then follows that the total number of such independent representations 
must equal the Hilbert-space dimension $4^{N_a^D}$. 
Here we show that for the Hubbard model on a bipartite lattice the number of independent representations of the group 
$SO(3)\times SO(3)\times U(1)$ is indeed $4^{N_a^D}$. In contrast, the number
of independent representations of the group $SO(4)$ is for that model found to be smaller
than its Hilbert-space dimension $4^{N_a^D}$. This is then consistent with the 
global symmetry of the Hubbard model on a bipartite lattice being larger
than $SO(4)$ and given by $SO(3)\times SO(3)\times U(1)$.

The rotated-electron occupancy configurations involving the (i) singly occupied
and (ii) unoccupied and doubly-occupied sites are
independent. They refer to the state representations of the spin $SU(2)$ symmetry 
$M_s=2S_c$ spin-$1/2$ spins and $\eta$-spin $SU(2)$ 
symmetry $M_{\eta}=2S_c^h$ 
$\eta$-spin-$1/2$ $\eta$-spins, respectively. Indeed, concerning the $\eta$-spin $SU(2)$ 
representations the rotated-electron doubly occupied sites and unoccupied sites
play the role of down and up $\eta$-spin-$1/2$ $\eta$-spins, respectively.
In turn, the $U(1)$ symmetry state representations refer 
to the relative occupancy configurations of the $2S_c$ 
rotated-electron singly-occupied sites and $2S_c^h$ rotated-electron unoccupied and doubly-occupied sites. 
For $U/4t\neq 0$, the Hilbert space can 
then be divided into a set of subspaces with fixed $S_{\eta}$, $S_s$, and 
$S_c$ values and thus with the same values 
$M_{\eta}=2S_c^h$ of $\eta$-spins and $M_{s}=2S_c$ of spins. 
The number of  $SU(2)\times SU(2)$ state representations with both
fixed values of $S_{\eta}$ and $S_s$,
which one can generate from $M_{\eta}$ 
$\eta$-spin-$1/2$ $\eta$-spins and $M_{s}$ spin-$1/2$ spins, 
reads ${\cal{N}}(S_{\eta} ,M_{\eta}).\,{\cal{N}}(S_s ,M_s)$. Here,
\begin{equation}
{\cal{N}} (S_{\alpha},M_{\alpha}) = (2S_{\alpha} +1)\left\{
{M_{\alpha}\choose M_{\alpha}/2-S_{\alpha}} - {M_{\alpha}\choose
M_{\alpha}/2-S_{\alpha}-1}\right\} \, , \label{Ncs}
\end{equation}
for $\alpha =\eta,s$. If for $U/4t\neq 0$ the global symmetry of the model 
was $SO(4)$, then the dimension of such a subspace would be 
${\cal{N}}(S_{\eta} ,M_{\eta}).\,{\cal{N}}(S_s ,M_s)$ and the sum of
all sub-space dimensions would give the Hilbert-space
dimension $4^{N_a^D}$. However, we confirm below that
such a sum obeys the inequality,
\begin{eqnarray}
& & \sum_{M_s=0}^{N_a^D}\,\sum_{S_{\eta}=0}^{M_{\eta}/2}\,
\sum_{S_s=0}^{M_s/2}\prod_{\alpha =\eta,s}{[1-(-1)^{[2S_{\alpha}-(M_{\eta}-M_s)/2]}]\over 2}
\,{\cal{N}}(S_{\alpha},M_{\alpha})
\nonumber \\
& = & \sum_{M_{\eta}=0}^{N_a^D}\,\sum_{S_{\eta}=0}^{M_{\eta}/2}\,
\sum_{S_s=0}^{M_s/2}\prod_{\alpha =\eta,s}{[1-(-1)^{[2S_{\alpha}-(M_{\eta}-M_s)/2]}]\over 2}
\,{\cal{N}}(S_{\alpha},M_{\alpha}) < 4^{N_a^D} \, ,
\label{Sc-0-Ntot}
\end{eqnarray}
and thus corresponds to a dimension smaller than $4^{N_a^D}$. Note that $M_{\eta}=[N_a^D-M_s]$
so that the numbers $M_{\eta}$ and $M_s$ are not independent. Therefore, the sums over
$M_s=0,1,...,N_a^D$ and $M_{\eta}=0,1,...,N_a^D$ are indeed alternative, as given in
Eq. (\ref{Sc-0-Ntot}).

In turn, that the model global symmetry is larger than $SO(4) = [SU(2)\times SU(2)]/Z_2$ and given
by $SO(3)\times SO(3)\times U(1)=[SU(2)\times SU(2)\times U(1)]/Z_2^2$ 
requires that one accounts for the number of representation states
of the extra global $U(1)$ symmetry. For $U/4t\neq 0$ it has in the subpaces considered here,
\begin{equation}
d_c = {N_a^D\choose 2S_c}= {N_a^D\choose 2S_c^h} \, ,
\label{d-c}
\end{equation}
representation states. Thus rather than ${\cal{N}}(S_{\eta} ,M_{\eta}).\,{\cal{N}}(S_s ,M_s)$
each of such subspaces has a larger dimension, 
\begin{equation}
d (S_{\eta},S_s,S_c) = d_c.\,{\cal{N}}(S_{\eta} ,N_a^D-2S_c).\,{\cal{N}}(S_s ,2S_c)
\, . \label{LCS}
\end{equation}
By performing the sum over all subspaces, one then finds indeed in Appendix A that,
\begin{widetext}
\begin{eqnarray}
{\cal{N}}_{tot} & = & 
\sum_{S_c=0}^{[N_a^D/2]}\,\sum_{S_{\eta}=0}^{[N_a^D/2-S_c]}\,
\sum_{S_s=0}^{S_c}{N_a^D\choose
2S_c} \prod_{\alpha =\eta,s}{[1+(-1)^{[2S_{\alpha}+2S_c]}]\over 2}
\,{\cal{N}}(S_{\alpha},M_{\alpha}) = \sum_{S_{\eta}=0}^{[N_a^D/2]}\,
\sum_{S_s=0}^{[N_a^D/2-S_{\eta}]}{[1+(-1)^{[2S_{\eta}+2S_s]}]\over 2}\nonumber \\
& \times & (2S_{\eta} +1)\, (2S_s +1)\,\Bigl[{N_a^D\choose N_a^D/2-S_{\eta}+S_s}\left\{{N_a^D\choose
N_a^D/2-S_{\eta}-S_s}+{N_a^D\choose N_a^D/2-S_{\eta}-S_s-2}\right\}\nonumber \\
& - & {N_a^D\choose N_a^D/2-S_{\eta}-S_s-1}\left\{{N_a^D\choose
N_a^D/2-S_{\eta}+S_s+1}+{N_a^D\choose N_a^D/2-S_{\eta}+S_s-1}\right\}\Bigr] = 4^{N_a^D} 
\, . \label{Ntot}
\end{eqnarray}
\end{widetext}

Finally, except that a factor of one in each term of the two 
alternative sums of Eq. (\ref{Sc-0-Ntot}) 
is replaced by the dimension $d_c$ in the sum of Eq. (\ref{Ntot}),
such sums are identical. This is confirmed by transforming the sums over
$M_s=0,1,...,N_a^D$ and $M_{\eta}=0,1,...,N_a^D$ of Eq. (\ref{Sc-0-Ntot}) in sums over
$S_c=M_s/2=0,1/2,...,N_a^D/2$ and $S_c=N_a^D/2-M_{\eta}/2=N_a^D/2, N_a^D/2-1/2,...,1/2,0$, respectively,
and accounting for that $2S_c$ can be written as $2S_c = N_a^D/2 - (M_{\eta}-M_s)/2$ where
$N_a^D/2$ is an odd integer number. That the dimension of Eq. (\ref{d-c}) obeys the
inequality $d_c\geq 1$ then implies the validity of the inequality
given in Eq. (\ref{Sc-0-Ntot}).

\subsection{Relation of the global symmetry to the exact solution of the 1D model}

In the particular case of the bipartite 1D lattice the Hubbard model has an exact solution
\cite{Lieb,Takahashi,Martins}. Since the global $SO(3)\times SO(3)\times U(1)$ symmetry
found here refers to 1D as well, it must be related to that exact solution. Such a solution refers to
the 1D Hubbard model in the subspace spanned by the highest-weight states (HWSs) or
lowest-weight states (LWSs) of both the $\eta$-spin $SU(2)$ and spin $SU(2)$ algebras. 
The model energy eigenstates that are HWSs or LWSs of these algebras are often 
called {\it Bethe states}. In order to clarify such a relation, rather than the so called coordinate 
Bethe ansatz \cite{Lieb,Takahashi}, it is convenient to consider 
the exact solution of the problem by the algebraic operator formulation of Ref. \cite{Martins}.
Within the latter formulation the HWSs or LWSs of the $\eta$-spin and spin algebras are built up in 
terms of linear combination of products of several types of
annihilation or creation fields acting onto the hole or electronic vacuum, 
respectively.  

The algebraic formulation of the Bethe states refers to the 
transfer matrix of the classical coupled spin model, which is the ``covering" 
1D Hubbard model \cite{CM}. Indeed, within the inverse scattering method 
\cite{Martins,ISM} the central object to be 
diagonalized is the quantum transfer matrix rather than the underlying 
1D Hubbard model. The transfer-matrix eigenvalues provide the spectrum 
of a set of conserved charges. The diagonalization of the charge degrees of freedom involves a transfer 
matrix associated with a charge monodromy 
matrix of the form provided in Eq. (21) of Ref.
\cite{Martins}. Its off-diagonal 
entries are some of the creation and annihilation fields. 
The commutation relations of such important operators are given in Eqs. 
(25), (40)-(42), (B.1)-(B.3), (B.7)-(B.11), and (B.19)-(B.22) of that reference. 
The solution of the spin degrees of freedom involves the 
diagonalization of the auxiliary transfer matrix associated with the spin monodromy 
matrix provided in Eq. (95) of Ref. \cite{Martins}. Again, the off-diagonal entries 
of that matrix play the role of creation and annihilation fields, whose 
commutation relations are given in Eq. (98) of that reference. 
The latter commutation relations correspond to the usual
Faddeev-Zamolodchikov algebra associated with the traditional
ABCD form of the elements of the monodromy matrix \cite{ISM}. It also applies to the
1D isotropic Heinsenberg model, whose global symmetry is $SU(2)$. 
Consistently, at half filling and for large $U/4t$ values the latter model describes the spin degrees of freedom
of the 1D Hubbard model. In turn, the above 
relations associated with the charge monodromy matrix refer to a 
different algebra. The corresponding form of that matrix is
called ABCDF by the authors of Ref. \cite{Martins}.

The main reason why the solution of the problem by the algebraic
inverse scattering method \cite{Martins} was achieved only thirty years
after that of the coordinate Bethe ansatz \cite{Lieb,Takahashi} is
that it was expected that the charge and spin monodromy 
matrices had the same traditional ABCD form, found previously for the related 1D isotropic Heinsenberg model \cite{ISM}. 
Indeed, such an expectation was that consistent with the occurrence
of a spin $SU(2)$ symmetry and a charge (and
$\eta$-spin) $SU(2)$ symmetry known long ago \cite{Zhang},
associated with a global $SO(4)=[SU(2)\times SU(2)]/Z_2$ 
symmetry. If that was the whole global symmetry of the 1D Hubbard model,
the charge and spin sectors would be associated with the
$\eta$-spin $SU(2)$ symmetry and spin $SU(2)$ symmetry,
respectively. A global $SO(4)=[SU(2)\times SU(2)]/Z_2$ symmetry
would then imply that the charge and spin monodromy 
matrices had indeed the same Faddeev-Zamolodchikov ABCD form.

However, all tentative schemes using charge and spin monodromy 
matrices of the same ABCD form failed to achieve the Bethe-ansatz
equations obtained by means of the coordinate Bethe ansatz \cite{Lieb,Takahashi}. 
Fortunately, the problem was solved by Martins and Ramos, who
used an appropriate representation of the charge and spin monodromy 
matrices, which allows for possible {\it hidden symmetries} \cite{Martins}.
Indeed, the structure of the charge and spin monodromy 
matrices introduced by these authors is able to distinguish creation and
annihilation fields as well as possible hidden symmetries.

Our results refer to the Hubbard model on any bipartite lattice.
Hence for the particular case of the bipartite 1D lattice they 
show that the hidden symmetry beyond $SO(4)$ is the charge global 
$U(1)$ symmetry found in this paper. Our studies reveal 
that for $U/4t>0$ the model charge and spin degrees of 
freedom are associated with $U(2)=SU(2)\times U(1)$ and
$SU(2)$ symmetries, rather than with two $SU(2)$ symmetries,
respectively. The occurrence of such charge $U(2)=SU(2)\times U(1)$
symmetry and spin $SU(2)$ symmetry is behind
the different ABCDF and ABCD forms of the charge and spin monodromy 
matrices of Eqs. (21) and (95) of Ref. \cite{Martins}, respectively.
Indeed, the former matrix is larger than the latter and involves 
more fields than expected from the model global
$SO(4)=[SU(2)\times SU(2)]/Z_2$ symmetry alone. This follows from
the global symmetry of the model on the 1D and other bipartite lattices 
being $SO(3)\times SO(3)\times U(1)=[SU(2)\times U(2)]/Z_2^2$ 
rather than $SO(4)=[SU(2)\times SU(2)]/Z_2$, as found in this paper. Hence our general results for the 
Hubbard model on a bipartite lattice are consistent with the
algebraic operator formulation of its exact solution for the particular
case of the 1D lattice \cite{Martins}.

\section{Concluding remarks}

On a square lattice, the Hubbard model is one of the most studied condensed-matter quantum problems.
Furthermore, on any bipartite lattice it is the simplest realistic toy model for description of the 
electronic correlation effects in general many-electron problems with short-range interaction.
Therefore, that the global symmetry of the Hubbard model on a bipartite lattice is larger than 
$SO(4)$ and given by $SO(3)\times SO(3)\times U(1)$ is an important exact result in its own right.
Furthermore, the new found global symmetry is expected to have important physical consequences. 

The studies of Ref. \cite{companion2} on the Hubbard model on the square lattice use
a description in terms of quantum objects related to the rotated electrons. The introduction
of such a description involves the global symmetry found in this paper and corresponding transformation
laws under a suitable electron - rotated-electron unitary transformation of the type considered
here and in Ref. \cite{Stein}. The spinless $c$ fermion, spin-$1/2$ spinon, and $\eta$-spin-$1/2$ 
$\eta$-spinon operators of such a description are a generalization to $U/4t>0$ of the 
$U/4t\gg 1$ ``quasicharge", spin, and ``pseudospin" operators of Ref. \cite{Stellan-06},
respectively. The former quantum objects emerge from a suitable electron - rotated-electron 
unitary transformation.Their operators have the same expressions in terms of rotated-electron
creation and annihilation operators as those of Ref. \cite{Stellan-06} in terms of electron
creation and annihilation operators, respectively. The occupancy configurations of the spinless $c$ fermions, spin-$1/2$ spinons, 
and $\eta$-spin-$1/2$ $\eta$-spinons generate a set of complete states. Those correspond 
to representations of the $U(1)$, spin $SU(2)$, and 
$\eta$-spin $SU(2)$ symmetries, respectively, associated with the three
dimensions of Eq. (\ref{LCS}) and the global symmetry found in this paper.

The square-lattice quantum liquid introduced in Ref. \cite{companion2} contains 
the one- and two-electron excitations of the Hubbard model on a square lattice.
At hole concentration $x=[N_a^2-N]/N_a^2=0$, $U/4t\approx 1.525$, and $t\approx 295$ meV it is found in that reference
to quantitatively describing the spin-wave spectrum observed in the parent compound 
La$_2$CuO$_4$ \cite{LCO-neutr-scatt}.
A system of weakly coupled planes, each described by the square-lattice quantum liquid of Ref. \cite{companion2},
is the simplest realistic toy model for the description of the role of correlations effects in the 
unusual properties of the cuprate hight-temperature superconductors \cite{ARPES-review,2D-MIT,duality}. 
After addition of such a weak three-dimensional
uniaxial anisotropy perturbation, the Hamiltonian terms that describe the fluctuations of two important pairing phases
are for intermediate $U/4t$ values found to have the same general form
as the microscopic Hamiltonian given in Eq. (1) of Ref. \cite{duality}. The main difference is that the 
electron creation and annihilation operators appear replaced by rotated-electron creation and
annihilation operators, respectively. Evidence is provided elsewhere that such a quantum liquid 
has for a well-defined hole-concentration range a long-range superconducting order. In addition, it seems 
indeed to contain some of the microscopic mechanisms behind the unusual properties of the
hole-doped cuprate hight-temperature superconductors.
It is commonly understood that Hamiltonian symmetries by themselves are not sufficient to prove 
that a particular symmetry is broken in the ground state. However, the symmetry of the action
that describes the fluctuations of the phases of such a quantum liquid and of that
of Ref. \cite{duality} is a global superconducting $U(1)$ symmetry. In the case of the former quantum liquid
the representations of such a $U(1)$ symmetry are generated by $c$ fermion occupancy configurations.
Thus it is directly related to the original model hidden global $U(1)$ symmetry found in this paper, whose
representations are also generated by $c$ fermion occupancy configurations. Such a preliminary 
result seems to confirm the important role plaid by the hidden $U(1)$ symmetry of the
global $SO(3)\times SO(3)\times U(1)$ symmetry found in this paper for the 
Hubbard model on a bipartite lattice. 

\begin{acknowledgments}
We thank Alejandro Muramatsu, Tiago C. Ribeiro, and Pedro D. Sacramento for illuminating discussions, 
Tobias Stauber for calling our attention to Ref. \cite{Stein} and for discussions, and the ESF Science programme 
INSTANS and the Portuguese PTDC/FIS/64926/2006 grant for support.
\end{acknowledgments}
\appendix

\section{Subspace-dimension summation} 

In this Appendix we perform the subspace-dimension summation of Eq. (\ref{Ntot}) that runs over $S_c$, $S_\eta$, and  $S_s$ 
integer and half-odd-integer values. For simplicity here we consider the square lattice so that $D=2$ in Eq. (\ref{Ntot}), yet the derivation 
proceeds in a similar way for any other $D$-dimensional cubic lattice where $D=1,2,3,...$. More generally, the sum-rule (\ref{Ntot})
is valid for the Hubbard model on any bipartite lattice. The subspace dimensions have the form
$d_r\cdot\prod_{\alpha=\eta,s}{\cal{N}}(S_{\alpha} ,M_{\alpha})$ given in Eq. (\ref{LCS})
where ${\cal{N}} (S_{\alpha},M_{\alpha})$ and $d_r$ are provided in Eqs (\ref{Ncs}) and (\ref{d-c}), respectively.
Recounting the terms of Eq. (\ref{Ntot}), one may choose $S_\eta$ to be the independent summation variable
what gives,
\begin{equation}\begin{split}
\sum_{S_c=0}^{N_a^2/2}\sum_{S_\eta=0}^{[N_a^2/2-S_c]}\sum_{S_s=0}^{S_c} &\frac{1+(-1)^{2(S_\eta + S_c)}}{2}  \;  \frac{1+(-1)^{2(S_s + S_c)}}{2}  \cdots = \\
 & = \sum_{S_\eta=0}^{N_a^2/2}\sum_{S_s=0}^{[\frac {N_a^2}{2}-S_\eta]}\sum_{S_c=S_s}^{[\frac {N_a^2}{2}-S_\eta]}\frac{1+(-1)^{2(S_\eta + S_s)}}{2} \; \frac{1+(-1)^{2(S_s + S_c)}}{2} \cdots \, .
\end{split}
\end{equation}
One can then rewrite the summation (\ref{Ntot}) in the form,
\begin{equation}
\mathcal{N}_{tot} = \sum_{S_\eta=0}^{N_a^2/2}\sum_{S_s=0}^{[N_a^2/2-S_\eta]}\frac{1+(-1)^{2(S_\eta + S_s)}}{2}(2S_\eta+1)(2S_s+1)\:\times \mathnormal{\Sigma}(S_\eta,S_s),
\label{Ntot2}
\end{equation}
where $\mathnormal{\Sigma}(S_\eta,S_s)$ denotes the $S_\eta$ and $S_s$ dependent summation over $S_c$
as follows,
\begin{eqnarray}
\mathnormal{\Sigma}(S_{\eta},S_s) & = &
\sum_{S_c=S_s}^{\frac {N_a^2}{2}-S_{\eta}} 
\frac{1+(-1)^{2(S_s + S_c)}}{2}  \binom{N_a^2}{2S_c} \times 
\nonumber \\ 
& & \Bigg[\binom{N_a^2-2S_c}{\frac {N_a^2}{2}-S_c-S_{\eta}} - \binom{N_a^2-2S_c}{\frac{N_a^2}{2}-S_c-S_{\eta}-1} \Bigg]
\Bigg[\binom{2S_c}{S_c-S_s}-\binom{2S_c}{S_c-S_s-1}\Bigg] 
\nonumber \\
&  & \sum_{S_c=S_s}^{\frac {N_a^2}{2}-S_\eta} 
\frac{1+(-1)^{2(S_s + S_c)}}{2}   N_a^2! \Bigg[\frac{1}{\displaystyle \left(S_c-S_s\right)! \left(S_c+S_s\right)!}-\frac{1}{\displaystyle \left( S_c-S_s-1\right)! \left( S_c+S_s+1\right)!}\Bigg] \times  
\nonumber \\ 
& & \Bigg[\frac{1}{\left( {N_a^2}/{2}-S_c-S_{\eta}\right)! \left( {N_a^2}/{2}-S_c+S_{\eta}\right)!} 
- \frac{1}{ \left( {N_a^2}/{2}-S_c-S_{\eta}-1\right))! \left( {N_a^2}/{2}-S_c+S_{\eta}+1\right)!}\Bigg] \, .
\label{sum-2}
\end{eqnarray}

In order to evaluate $\mathnormal{\Sigma}(S_\eta,S_s)$ it is useful to replace the variable $S_c$ by $k=S_c-S_s$.
To simplify the notation we then introduce,
\begin{equation}
\mathcal S = S_{\eta}+S_s = \mathcal S(S_\eta,S_s)
\, ; \hspace{0.35cm}
\mathcal D = S_{\eta}-S_s = \mathcal D(S_\eta,S_s) \, .
\label{SD}
\end{equation} 
Due to the parity factor, in the summation over $k$ only the terms with $k$ integer survive so that,
\begin{eqnarray}
\mathnormal{\Sigma} 
 & = &
\displaystyle \sum_{k=0 }^{\frac {N_a^2}{2}-\mathcal S}
N_a^2! \left[ \frac{1}{k! \left(\mathcal {S-D}+k\right)!} - \frac{1}{(k-1)! \left(\mathcal {S-D}+k+1\right)!} \right] \times
\nonumber \\
&& \hspace{1.5cm} \left[ \frac{1}{\left({N_a^2}/{2}-\mathcal S-k\right)! \left({N_a^2}/{2}+\mathcal D-k\right)!} - 
\frac{1}{ \left( {N_a^2}/{2}-\mathcal S-k-1\right)! \left({N_a^2}/{2}+\mathcal D-k+1\right)!} \right]
\nonumber \\
 & = &
\displaystyle \sum_{k=0}^{\frac {N_a^2}{2}-\mathcal S} 
N_a^2!  \Bigg \{ \frac{1}{k! \left(\mathcal {S-D}+k\right)!}\;\frac{1}{\left({N_a^2}/{2}-\mathcal S-k\right)! \left({N_a^2}/{2}+\mathcal D-k\right)!} \; -
\nonumber \\
&& \hspace{1.cm} - \;  
\frac{1}{k! \left(\mathcal {S-D}+k\right)!} \; \frac{1}{ \left( {N_a^2}/{2}-\mathcal S-k-1\right)! \left({N_a^2}/{2}+\mathcal D-k+1\right)!}   \; -
\nonumber \\
&& \hspace{1.cm} - \;  
\frac{1}{(k-1)! \left(\mathcal {S-D}+k+1\right)!} \; \frac{1}{\left({N_a^2}/{2}-\mathcal S-k\right)! \left({N_a^2}/{2}+\mathcal D-k\right)!}   \; +
\nonumber \\
&& \hspace{1.cm} + \; 
\frac{1}{(k-1)! \left(\mathcal {S-D}+k+1\right)!} \; \frac{1}{ \left( {N_a^2}/{2}-\mathcal S-k-1\right)! \left({N_a^2}/{2}+\mathcal D-k+1\right)!} \Bigg \} \, ,
\label{sum-3}
\end{eqnarray}
where now the $k$ summation runs over integers only.

In order to perform the summation (\ref{sum-3}) we rearrange the terms as follows,
\begin{eqnarray}
\mathnormal{\Sigma}
& = &  
\displaystyle \sum_{k=0}^{\frac {N_a^2}{2}-\mathcal S} 
\Bigg \{   \frac{1}{\left( {N_a^2}/{2}+\mathcal {D}\right)!\left( {N_a^2}/{2}-\mathcal {D}\right)!} 
\Bigg [\binom {N_a^2/2+\mathcal {D}}{k} \binom {N_a^2/2-\mathcal {D}}{N_a^2/2-\mathcal S-k} 
+\binom {N_a^2/2+\mathcal {D}}{k-1} \binom {N_a^2/2-\mathcal {D}}{N_a^2/2-\mathcal S-k-1} \Bigg ] -
\nonumber \\
&& - \; \frac{1}{\left( {N_a^2}/{2}+\mathcal {D}+1\right)! \left( {N_a^2}/{2}-\mathcal {D}-1\right)!} 
\binom {N_a^2/2+\mathcal {D}+1}{k} \binom {N_a^2/2-\mathcal {D}-1}{N_a^2/2-\mathcal S-k-1} \; -
\nonumber \\ 
&& - \; \frac{1}{\left( {N_a^2}/{2}+\mathcal {D}-1\right)! \left( {N_a^2}/{2}-\mathcal {D}+1\right)!} 
\binom {N_a^2/2+\mathcal {D}-1}{k-1} \binom {N_a^2/2-\mathcal {D}+1}{N_a^2/2-\mathcal S-k} \Bigg \}  N_a^2!   \, ,
\label{ARF2}
\end{eqnarray}
or
\begin{eqnarray}
\mathnormal{\Sigma}
& = &  
\displaystyle \binom {N_a^2}{N_a^2/2-\mathcal {D}} \sum_{k=0}^{\frac {N_a^2}{2}-\mathcal S} 
\Bigg [ \binom {N_a^2/2+\mathcal {D}}{k} \binom {N_a^2/2-\mathcal {D}}{N_a^2/2-\mathcal S-k} + 
\binom {N_a^2/2+\mathcal {D}}{k-1} \binom {N_a^2/2-\mathcal {D}}{N_a^2/2-\mathcal S-k-1} \Bigg ] \; -
\nonumber \\
&&  - \displaystyle \binom {N_a^2}{N_a^2/2-\mathcal {D}-1}  \sum_{k=0}^{\frac {N_a^2}{2}-\mathcal S}\binom {N_a^2/2+\mathcal {D}+1}{k} \binom {N_a^2/2-\mathcal {D}-1}{N_a^2/2-\mathcal S-k-1} \; -
\nonumber \\
&&  - \displaystyle \binom {N_a^2}{N_a^2/2-\mathcal {D}+1}  \sum_{k=0}^{\frac {N_a^2}{2}-\mathcal S} \binom {N_a^2/2+\mathcal {D}-1}{k-1} \binom {N_a^2/2-\mathcal {D}+1}{N_a^2/2-\mathcal S-k} \, .
\label{ARF}
\end{eqnarray}

Next, by using the identity,
\begin{equation}
\sum_{k=0}^{N} \binom {A}{k} \binom {B}{N-k}=\binom {A+B}{N} \, ,
\label{MJ}
\end{equation}
we carry out separately the summations in expression (\ref{ARF}), what gives,
\begin{equation}
\sum_{k=0}^{{N_a^2}/{2}-\mathcal S} \binom {N_a^2/2+\mathcal {D}}{k} \binom {N_a^2/2-\mathcal {D}}{N_a^2/2-\mathcal S-k} = \binom {N_a^2}{N_a^2/2-\mathcal S} \, ,
\label{MJ1}
\end{equation}
\begin{eqnarray}
\sum_{k=0}^{{N_a^2}/{2}-\mathcal S}\binom {N_a^2/2+\mathcal {D}}{k-1} \binom {N_a^2/2-\mathcal {D}}{N_a^2/2-\mathcal S-k-1}
& = &
\sum_{k=1}^{{N_a^2}/{2}-\mathcal S -1}\binom {N_a^2/2+\mathcal {D}}{k-1} \binom {N_a^2/2-\mathcal {D}}{N_a^2/2-\mathcal S-k-1}
\nonumber \\
& = &
\sum_{k'=0}^{ {N_a^2}/{2}-\mathcal S-2} \binom {N_a^2/2+\mathcal {D}}{k'} \binom {N_a^2/2-\mathcal {D}}{N_a^2/2-\mathcal S-2-k'} 
\nonumber \\
& = &
\binom {N_a^2}{N_a^2/2-\mathcal S-2} \, ,
\label{MJ2}
\end{eqnarray}
\begin{eqnarray}
\sum_{k=0}^{{N_a^2}/{2}-\mathcal S}\binom {N_a^2/2+\mathcal {D}+1}{k} \binom {N_a^2/2-\mathcal {D}-1}{N_a^2/2-\mathcal S-k-1} 
& = & 
\sum_{k=0}^{{N_a^2}/{2}-\mathcal S -1} \binom {N_a^2/2+\mathcal {D}+1}{k} \binom {N_a^2/2-\mathcal {D}-1}{N_a^2/2-\mathcal S-1-k}
\nonumber \\
& = &
\binom {N_a^2}{N_a^2/2-\mathcal S-1} \, ,
\label{MJ3}
\end{eqnarray}
and
\begin{eqnarray}
\sum_{k=0}^{{N_a^2}/{2}-\mathcal S}\binom {N_a^2/2+\mathcal {D}-1}{k-1} \binom {N_a^2/2-\mathcal {D}+1}{N_a^2/2-\mathcal S-k} 
& = & 
\sum_{k=1}^{{N_a^2}/{2}-\mathcal S}\binom {N_a^2/2+\mathcal {D}-1}{k-1} \binom {N_a^2/2-\mathcal {D}+1}{N_a^2/2-\mathcal S-k} 
\nonumber \\
& = &
\sum_{k'=0}^{{N_a^2}/{2}-\mathcal S-1}\binom {N_a^2/2+\mathcal {D}-1}{k'} \binom {N_a^2/2-\mathcal {D}+1}{N_a^2/2-\mathcal S-1-k'} 
\nonumber \\
& = &
\binom {N_a^2}{N_a^2/2-\mathcal S-1} \, .
\label{MJ4}
\end{eqnarray}

Introducing these results in expression (\ref{ARF}) for $\mathnormal{\Sigma}$ leads to,
\begin{eqnarray}
\mathnormal{\Sigma}(S_\eta,S_s)
& = & 
\displaystyle \binom {N_a^2}{N_a^2/2-\mathcal {D}} \left [\binom {N_a^2}{N_a^2/2-\mathcal S} + \binom {N_a^2}{N_a^2/2-\mathcal S-2} \right ] -
\nonumber \\
& & 
- \binom {N_a^2}{N_a^2/2-\mathcal S-1}   \left [ \displaystyle \binom {N_a^2}{N_a^2/2-\mathcal {D}+1} +  \displaystyle \binom {N_a^2}{N_a^2/2-\mathcal {D}-1}\right ]
\nonumber \\
&  \equiv  &  
\mathbf{\Sigma}(\mathcal S,\mathcal {D}) \, .
\label{sigma}
\end{eqnarray}

Expression (\ref {Ntot2}) for $\mathcal{N}_{tot}$ can now be rewritten as,
\begin{eqnarray}
\mathcal{N}_{tot} 
& = & 
\sum_{S_\eta=0}^{N_a^2/2}\sum_{S_s=0}^{[N_a^2/2-S_\eta]}\frac{1+(-1)^{2(S_\eta + S_s)}}{2}(2S_\eta+1)(2S_s+1) \times 
\nonumber \\
& & 
\Bigg \{ \displaystyle \binom {N_a^2}{N_a^2/2-(S_\eta-S_s)} \Bigg [\binom {N_a^2}{N_a^2/2-(S_\eta + S_s)} + \binom {N_a^2}{N_a^2/2-(S_\eta + S_s)-2} \Bigg ] -
\nonumber \\
& & 
- \binom {N_a^2}{N_a^2/2-(S_\eta + S_s)-1}   \Bigg [ \displaystyle \binom {N_a^2}{N_a^2/2-(S_\eta-S_s)+1} 
+  \displaystyle \binom {N_a^2}{N_a^2/2-(S_\eta-S_s)-1}\Bigg ] \Bigg \} \, ,
\label{N-tot-2}
\end{eqnarray}
where the summations run over both integers and half-odd integers. The use of the notation (\ref{SD}) then allows 
rewriting (\ref{N-tot-2}) in compact form,
\begin{equation}
\mathcal{N}_{tot} = \sum_{S_\eta=0}^{N_a^2/2}\sum_{S_s=0}^{[N_a^2/2-S_\eta]}\frac{1+(-1)^{2 \mathcal S}}{2}(\mathcal S + \mathcal {D}+1)(\mathcal S - \mathcal {D}+1)\:\times \mathbf{\Sigma}(\mathcal S,\mathcal {D}) \, ,
\label{Ntot*}
\end{equation}
where the summations run again over both integers and half-odd integers.

We can perform the summations of Eq. (\ref{Ntot*}) in the integers $\mathcal S$ and $ \mathcal {D}$ instead of in $S_\eta$  and $S_s$. 
Indeed, the first factor cancels all the terms with $\mathcal S$ and $ \mathcal {D}$ non-integer so that,
\begin{equation}\begin{split}
\sum_{S_\eta=0}^{N_a^2/2}\sum_{S_s=0}^{[N_a^2/2-S_\eta]}\frac{1+(-1)^{2 ( S_\eta+  S_s)}}{2} \cdots &  
\mbox {($S_\eta$ and $S_s$ both either integers or half odd integers)} = \\
& \qquad \qquad =\sum_{\mathcal S=0}^{N_a^2/2}\sum_{\mathcal {D}=-\mathcal S}^{+\mathcal S} \cdots \mbox {($\mathcal S$ 
and $\mathcal {D}$ integers)} \, .
\nonumber
\end{split}\end{equation}

Thus we find,
\begin{equation}
\mathcal{N}_{tot} = \sum_{\mathcal S=0}^{N_a^2/2}\sum_{\mathcal {D}=-\mathcal S}^{+\mathcal S} \left((\mathcal S +1)^2 - \mathcal {D}^2\right) \:\times \mathbf{\Sigma}(\mathcal S,\mathcal {D}) \, .
\nonumber
\end{equation}
The use of the result (\ref{sigma}) then leads to,
\begin{equation}\begin{split}
\mathcal{N}_{tot} = \sum_{\mathcal S=0}^{N_a^2/2}\sum_{\mathcal {D}=-\mathcal S}^{\mathcal S} &\left((\mathcal S +1)^2 - \mathcal {D}^2\right) \Bigg \{ \displaystyle \binom {N_a^2}{N_a^2/2-\mathcal {D}} \left [\binom {N_a^2}{N_a^2/2-\mathcal S} + \binom {N_a^2}{N_a^2/2-\mathcal S-2} \right ] -
 \\
 & \qquad \qquad - \binom {N_a^2}{N_a^2/2-\mathcal S-1}   \left [ \displaystyle \binom {N_a^2}{N_a^2/2-\mathcal {D}+1} +  \displaystyle \binom {N_a^2}{N_a^2/2-\mathcal {D}-1}\right ]\Bigg \} \, .
\end{split}\label{NtotSD}
\end{equation}

Replacing the variable $\mathcal S$ by $\mathcal S' = \mathcal S +1$ we reach a more tractable expression for $\mathcal{N}_{tot}$,
\begin{equation}
\mathcal{N}_{tot} = \sum_{\mathcal S'=1}^{N_a^2/2+1}\sum_{\mathcal {D}=-\mathcal S'+1}^{\mathcal S'-1} \mathcal T(S',D) \, ,
\label{NtotS'D2}
\end{equation}
where
\begin{equation}\begin{split}
\mathcal {T(S',D)} &=  \left( (\mathcal S')^2 - \mathcal {D}^2\right)\:\times \mathbf{\Sigma}(\mathcal S'-1,\mathcal {D}) \\
 &= \left(\mathcal S'^2 - \mathcal {D}^2\right) \Bigg \{\binom {N_a^2}{N_a^2/2-\mathcal {D}}\left [\binom {N_a^2}{N_a^2/2-\mathcal S'+1} + \binom {N_a^2}{N_a^2/2-\mathcal S'-1} \right ] - \\
  & \hspace{2.2cm}  - \binom {N_a^2}{N_a^2/2-\mathcal S'}   \left [ \displaystyle \binom {N_a^2}{N_a^2/2-\mathcal {D}+1} +  \displaystyle \binom {N_a^2}{N_a^2/2-\mathcal {D}-1}\right ] \Bigg \} \, ,
\end{split}\label{TermS'D}
\end{equation}
is completely symmetric in the summation variables.

Since $\mathcal {T(S',D=\pm S')}=0$, we can extend the summation over $\mathcal {D}$ of Eq.(\ref{NtotS'D2}) to $\mathcal {D} = \pm \mathcal S'$. 
We then formally extend the summation over $ \mathcal S' $ to $ \mathcal S'= 0 $ because the corresponding term vanishes:  
$\mathcal T(\mathcal S'=0,\mathcal D=0)=0 $. Futhermore, $\mathcal {T(\pm S',D)}=\mathcal {T(S',\pm D)}=\mathcal {T(S',D)}$, and  due to the symmetry 
$\mathcal S' \leftrightarrow \mathcal {D}$ we can write,
\begin{equation}
\sum_{\mathcal S'=1}^{N_a^2/2}\sum_{\mathcal {D}=-\mathcal S'+1}^{\mathcal S'-1} \mathcal {T(S',D)} = \frac{1}{4}\sum_{\mathcal S',\mathcal {D}=-(N_a^2/2+1)}^{N_a^2/2+1} \mathcal {T(S',D)} \, .
\label{ah}
\end{equation}

Let us introduce the numbers $p$ and $q$ such that,
\begin{eqnarray*}
\mathcal S'+N_a^2/2+1=p &\Leftrightarrow &  \mathcal S'=p-(N_a^2/2+1)   \\ 
\mathcal {D}+N_a^2/2+1=q &\Leftrightarrow & \mathcal {D}=q-(N_a^2/2+1) \, .
\label{pq}
\end{eqnarray*}
The use of (\ref{ah}) then allows rewriting (\ref{NtotS'D2}) as,
\begin{equation}\begin{split}
\mathcal{N}_{tot}  = \frac{1}{4}\sum_{p,q=0}^{N_a^2+2} & \left [ q(N_a^2+2-q)-p(N_a^2+2-p) \right ] \\
& \times \left \{ \binom {N_a^2}{q-1}\left [\binom {N_a^2}{p}+\binom {N_a^2}{p-2}\right ]- \binom {N_a^2}{p-1}\left [\binom {N_a^2}{q}+\binom {N_a^2}{q-2}\right ]\right \} \, .
\end{split}
\label {uff}
\end{equation}

This expression can be simplified noticing that,
\begin{equation}
\binom {N}{x} + \binom {N}{x-2} = -2 \binom {N}{x-1} + \binom {N+2}{x} \, .
\nonumber
\end{equation}
Replacing in Eq.(\ref {uff}) one then finds,
\begin{equation}\begin{split}
\mathcal{N}_{tot} 
& =   \frac{1}{4}\sum_{p,q=0}^{N_a^2+2} \Bigg [ q(N_a^2+2-q)-p(N_a^2+2-p) \Bigg ] \left \{ \binom {N_a^2}{q-1}\binom {N_a^2+2}{p}- \binom {N_a^2}{p-1} \binom {N_a^2+2}{q}\right \}  \\
& = \frac{1}{4}\sum_{p,q=0}^{N_a^2+2} \left \{q(N_a^2+2-q)\left [ \binom {N_a^2}{q-1}\binom {N_a^2+2}{p}- \binom {N_a^2}{p-1} \binom {N_a^2+2}{q}\right ] + (q \leftrightarrow p)\right \}
\\
& =  \frac{1}{4} 2  \Bigg \{ \; \sum_{q=0}^{N_a^2+2} q(N_a^2+2-q) \binom {N_a^2}{q-1} \sum_{p=0}^{N_a^2+2} \binom {N_a^2+2}{p} -  \sum_{q=0}^{N_a^2+2} q(N_a^2+2-q)\binom {N_a^2+2}{q} \sum_{p=0}^{N_a^2+2} \binom {N_a^2}{p-1} \Bigg \} \, .
\end{split}\end{equation}

Finally, the use of the identities,
\begin{equation}
\sum_{k=0}^{N} \binom {N}{k} = 2^{N} \, ,
\nonumber
\end{equation}
\begin{equation}
\sum_{k=0}^{N+2} \binom {N}{k-1} =\sum_{k=1}^{N+1} \binom {N}{k-1}  =\sum_{k'=0}^{N} \binom {N}{k'} = 2^N \, ,
\nonumber
\end{equation}
\begin{eqnarray*}
\sum_{k=0}^{N} k(N-k)\binom {N}{k}  &=& \sum_{k=1}^{N-1} \frac {N!}{(k-1)!(N-k-1)!}=  N(N-1) \sum_{k-1=0}^{N-2} \binom {N-2}{k-1} \\
& =& N(N-1) 2^{N-2} \, ,
\nonumber
\end{eqnarray*}
and
\begin{eqnarray*}
\sum_{k=0}^{N+2} k(N+2-k) \binom {N}{k-1} 
& = & 
\sum_{k=1}^{N+1}  k(N+2-k)\binom {N}{k-1} = \sum_{k-1=0}^{N} k(N+2-k)\binom {N}{k-1} \\ 
& = & \sum_{k'=0}^{N} (k'+1)(N-k'+1)\binom {N}{k'} = \sum_{k'=0}^{N} \left[ k'(N-k') + (N+1) \right ]\binom {N}{k'} \\
& = & N(N-1) 2^{N-2} +(N+1) 2^N = 2^{N-2} \left[  N(N-1)+4(N+1)\right ] \\
& = & \left[  N^2 + 3N + 4 \right ]2^{N-2} \, ,
\end{eqnarray*}
leads to,
\begin{equation}
\mathcal{N}_{tot} =  \frac{1}{2}  \left \{  \left[  N_a^2 + 3N_a^2 + 4 \right ]2^{N_a^2-2} \times 2^{N_a^2+2}  -  (N_a^2+2)(N_a^2+1) 2^{N_a^2} \times 2^{N_a^2}\right \} = \frac{1}{2}2^{2N_a^2} \times 2  = 4^{N_a^2} \, ,
\end{equation}
which is the desired result.


\begin{references}
\bibitem{Zoller}
        D. Jaksch, P. Zoller, Ann. Phys. 315 (2005) 52.   
\bibitem{cubic}
	R. J\"ordens, N. Strohmaier, K. G\"unter, H. Moritz, T. Esslinger,  
	Nature 455 (2008) 204.
\bibitem{Claessen}
	M. Sing, U. Schwingenschl\"ogl, R. Claessen, P.
        Blaha, J. M. P. Carmelo, L. M. Martelo, P. D. Sacramento,
        M. Dressel, C. S. Jacobsen,
        Phys. Rev. B 68 (2003) 125111.	
\bibitem{TTF-TCNQ}    
        J. M. P. Carmelo, D. Bozi, K. Penc,
        J. Phys.: Cond. Matt. 20 (2008) 415103;
        D. Bozi, J. M. P. Carmelo, K. Penc, P. D. Sacramento, J.
 	Phys.: Cond. Matt. 20 (2008) 022205.        
\bibitem{ARPES-review} 
	A. Damascelli, Z. Hussain, Z.-X. Shen, 
	Rev. Mod. Phys. 75 (2003) 473.
\bibitem{2D-MIT} 
	P. A. Lee, N. Nagaosa, X.-G. Wen, 
	Rev. Mod. Phys. 78 (2006) 17.
\bibitem{duality}
	Z. Te${\rm\check{s}}$anovi\'c,   
	Nature Phys., 4 (2008) 408.
\bibitem{Lieb} 
	E. H. Lieb, F. Y. Wu, 
	Phys. Rev. Lett. 20 (1968) 1445.
\bibitem{Takahashi}
	Minoru Takahashi, Progr. Theor. Phys 47 (1972) 69.  
\bibitem{Martins} 
	M. J. Martins, P. B. Ramos, Nucl. Phys. B 522 (1998) 413.
 \bibitem{Zhang}
 	O. J. Heilmann, E. H. Lieb, Ann. N. Y. Acad. Sci.
	172 (1971) 583;
	E. H. Lieb, Phys. Rev. Lett. 62 (1989) 1201;  
	C. N. Yang, S. C. Zhang, Mod. Phys. Lett. B 4 (1990) 759;       
	S. C. Zhang, Phys. Rev. Lett. 65 (1990) 120. 
\bibitem{U(1)-NL} 
	Stellan \" Ostlund, Eugene Mele, 
	Phys. Rev. B 44 (1991) 12413.
\bibitem{Stein} 
	J. Stein, J. Stat. Phys. 88 (1997) 487.
\bibitem{Stellan-06}         
         Stellan \"Ostlund, Mats Granath, Phys. Rev. Lett. 96 (2006) 066404.  
\bibitem{CM}
        B. Sriram Shastry, J. Stat. Phys. 50 (1988) 57.              
\bibitem{ISM}
        E. K. Sklyanin, L. A. Takhtadzhan, L. D.
        Faddeev, Theor. Math. Fiz. 40 (1979) 194.	
\bibitem{companion2}
	J. M. P. Carmelo, Nucl. Phys. B 824 (2010) 452 and references therein.
\bibitem{LCO-neutr-scatt} 
	R. Coldea, S. M. Hayden, G. Aeppli, T. G. Perring, C. D. Frost, 
	T. E. Mason, S.-W. Cheong, Z. Fisk,
	Phys. Rev. Lett. 86 (2001) 5377.
\end{references}
\end{document}